\documentclass[12pt]{article}

\usepackage{psfig}

\usepackage[dvips]{color}

\newcommand{\Black}{\color [rgb]{0,0,0}}

\newcommand{\Brown}{\color [rgb]{0.4,0.1,0.1}}

\def\TL{\hfil$\displaystyle{##}$}
\def\TR{$\displaystyle{{}##}$\hfil}

\def\TT{\hbox{##}}
\def\seqalign#1#2{\vcenter{\openup1\jot
  \halign{\strut #1\cr #2 \cr}}}

\def\fixit#1{}

\def\mop#1{\mathop{\rm #1}\nolimits}

\def\Vol{\mop{Vol}}

\def\overleftrightarrow#1{\vbox{\ialign{##\crcr
     $\leftrightarrow$\crcr\noalign{\kern-0pt\nointerlineskip}
     $\hfil\displaystyle{#1}\hfil$\crcr}}}

\def\lsim{\mathrel{\mathstrut\smash{\ooalign{\raise2.5pt\hbox{$<$}\cr\lower2.5pt\hbox{$\sim$}}}}}
\def\gsim{\mathrel{\mathstrut\smash{\ooalign{\raise2.5pt\hbox{$>$}\cr\lower2.5pt\hbox{$\sim$}}}}}

\def\sqr#1#2{{\vcenter{\vbox{\hrule height.#2pt
         \hbox{\vrule width.#2pt height#1pt \kern#1pt
            \vrule width.#2pt}
         \hrule height.#2pt}}}}

\def\href#1#2{#2}

\def\lbldef#1#2{\expandafter\gdef\csname #1\endcsname {#2}}
\def\eqn#1#2{\lbldef{#1}{(\ref{#1})}%
\begin{equation} \eqalign{#2} \label{#1} \end{equation}}
\def\eqalign#1{\vcenter{\openup1\jot
    \halign{\strut\span\TL & \span\TR\cr #1 \cr
   }}}

\begin{document}
\pagestyle{plain}
\setcounter{page}{1}
\begin{titlepage}

\begin{flushright}
PUPT-2059 \\
hep-th/0301002
\end{flushright}
\vfil

\begin{center}
{\huge Universality classes for horizon instabilities}
\end{center}

\vfil
\begin{center}
{\large Steven S. Gubser$^1$ and Arkadas Ozakin$^2$}
\end{center}

$$\seqalign{\span\TL & \span\TT}{
^1 & Joseph Henry Laboratories, Princeton University, Princeton, NJ 08544 \cr
^2 & Lauritsen Laboratories of Physics, Caltech, Pasadena, CA  91125
}$$
\vfil

\begin{center}
{\large Abstract}
\end{center}

\noindent
 We introduce a notion of universality classes for the Gregory-Laflamme
instability and determine, in the supergravity approximation, the
stability of a variety of solutions, including the non-extremal
D3-brane, M2-brane, and M5-brane.  These three non-dilatonic branes
cross over from instability to stability at a certain non-extremal
mass.  Numerical analysis suggests that the wavelength of the shortest
unstable mode diverges as one approaches the cross-over point from
above, with a simple critical exponent which is the same in all three
cases.

\vfil
\begin{flushleft}
December 2002
\end{flushleft}
\end{titlepage}
\newpage
\tableofcontents

\section{Introduction}
\label{Introduction}

The Gregory-Laflamme instability \cite{glOne,glTwo} has been
conjectured to arise precisely when local thermodynamic instabilities
exist, for horizons with infinite extent and a translational symmetry
\cite{gmOne,gmTwo}.  The idea is that, barring finite volume effects
or some unexpected consequences of a curved horizon, the horizon will
tend to become lumpy through real-time dynamics precisely if it can
gain entropy by doing so.  Based on this line of thinking, one might
poetically ascribe the dynamical stability of the Schwarzschild black
hole in four dimensions merely to the fact that the horizon is smaller
in total size than the shortest wavelength instability.

A fairly thorough analysis in \cite{Reall}, exploiting a connection
between time-independent, finite-wavelength perturbations in
Lorentzian signature and negative modes in Euclidean signature, gives
considerable confidence that the conjecture of \cite{gmOne,gmTwo} is
correct, as well as making clearer the reasons why thermodynamic and
dynamical stability should be connected.  The analysis of \cite{Reall}
does leave several points open, including the following:
 \begin{enumerate}
  \item For the black brane solutions familiar from string theory, just
when is stability lost?
  \item What happens near the threshold of stability?
  \item Once an instability develops, what does the system evolve
into?
 \end{enumerate}
Question 3) has been the subject of recent scrutiny for the case of
uncharged black branes (see for instance
\cite{hm,gBead,kol,wiseman,wisemanTwo}).  Question 1) is straightforward
if one grants the analysis of \cite{Reall}, since one needs only
compute the specific heat as a function of non-extremality.  We will
address the question in section~\ref{thermo} from this thermodynamic
point of view, and in section~\ref{Cross} by directly searching for
time-independent, finite-wavelength perturbations, thus providing an
independent check on the validity of the analysis of \cite{Reall}.  We
will also make a start on question 2), based on our numerics.  In
section~\ref{universality} we will introduce a notion of universality
classes for black brane horizons, based on solutions described in
\cite{cr}.  For the questions we will address, a single characteristic
exponent defines a universality class: this exponent describes how
horizon entropy scales with temperature.

Our notion of universality classes is of some intuitive use in
understanding the stability properties of various charged brane
solutions.  In particular, the well-known non-extremal D$p$-branes of
type II string theory interpolate between one universality class far
from extremality (where the entropy scales as a negative power of
temperature, and Gregory-Laflamme instabilities exist) to a different
universality class near extremality (where entropy scales as a
positive power of temperature, and Gregory-Laflamme instabilities
probably do not exist).

Other recent work on the Gregory-Laflamme instability includes
\cite{Hirayama,harmark,hubeny}.  In particular, \cite{Hirayama}
explores the stability of charged brane solutions, using methods
somewhat different from ours, but with conclusions that overlap and
agree with those in section~\ref{Cross}.

\section{Thermodynamic considerations}
\label{thermo}

The D$p$-brane solutions to type~II string theory have the following
string frame metric and dilaton:
  \eqn{SMetric}{
   ds^2 &= G_{tt} dt^2 + G_{rr} dr^2 + G_{\theta\theta} d\Omega_{n+1}^2 +
            G_{yy} dy^i dy^i \cr
        &= -{h \over \sqrt{f}} dt^2 + {\sqrt{f} \over h} dr^2 +
           \sqrt{f} r^2 d\Omega_{n+1}^2 +
           {1 \over \sqrt{f}} dy^i dy^i \cr
   e^{2 \phi} &= f^{(n-4)/2}
  }
where
  \eqn{HandF}{
   h = 1 - {r_0^n \over r^n} \qquad\qquad
   f = 1 + {r_0^n \sinh^2 \alpha \over r^n} \,.
  }
The string metric $G_{\mu\nu}$ is related to the Einstein metric
$g_{\mu\nu}$ by $g_{\mu\nu} = e^{-\phi/2} G_{\mu\nu}$.  In
equations~\SMetric\ and \HandF, $n=7-p$.  We will assume throughout
that the metric \SMetric\ is being used to describe a large number
$N_p$ of coincident D$p$-branes, so that the supergravity
approximation is reliable (except perhaps near the horizon at
extremality, where some of the solutions have a null singularity).

The thermodynamic properties of D$p$-branes are well-known, so we will
present here only a brief summary.  The original literature, including
\cite{hs,hp,imsy}, can be consulted for further details.  The entropy
of the D$p$-brane solution \SMetric\ is
  \eqn{BekHawk}{
   S = {A_{\rm Einstein} \over 4 G_N} =
    {A_{\rm string} \over 4 G_N e^{2\phi}} =
    {4 \pi \Omega_{n+1} \over 2 \kappa^2} V r_0^{n+1} \cosh\alpha \,,
  }
where $A_{\rm Einstein}$ is the horizon area measured using the
Einstein frame metric, $A_{\rm string}$ is the horizon area in the
string frame metric, $V$ is the coordinate volume in the $y^i$
directions, and $\Omega_{n+1}$ is the volume of the sphere $S^{n+1}$:
 \eqn{SphereVol}{
  \Omega_m = \Vol S^m = 2 {\pi^{m+1 \over 2} \over
   \Gamma\left( {m+1 \over 2} \right)} \,.
 }
If one compactified the $y^i$ on a torus whose volume at infinity was
$V$, then \BekHawk\ would be the total entropy of the resulting
horizon.  The ADM mass (most easily calculated using the Einstein
frame metric) is
  \eqn{ADMMass}{
   M = {\Omega_{n+1} V \over 2 \kappa^2} r_0^n
    \left( {n+2 \over 2} + {n \over 2} \cosh 2\alpha \right) \,.
  }
The temperature can be extracted from the surface gravity:
 \eqn{Tcl}{
  T = {\kappa \over 2 \pi} = {1 \over 2\pi}
   \left. {\partial g_{tt} / \partial r \over 2 \sqrt{g_{tt} g_{rr}}}
    \right|_{r=r_0} = {n \over 4 \pi r_0 \cosh\alpha} \,.
 }
One can substitute $G_{tt}$ and $G_{rr}$ for $g_{tt}$ and $g_{rr}$ in
\Tcl, provided the dilaton is non-singular at the horizon (which would
just mean that the horizon is a null singularity, and that only
happens for extremal D$p$-branes).

It is straightforward to check that $T = \partial M / \partial S$.

The D$p$-brane charge is quantized, and can be expressed in terms of the
following constraint:
  \eqn{QuantRA}{
   \tilde{R}^n \equiv {\textstyle{1 \over 2}} r_0^n \sinh 2\alpha =
    {2 \kappa^2 \over n \Omega_{n+1}} N_p \tau_p
    = {N_p \sqrt{ 2\kappa^2 (2 \pi)^{7-2p} (\alpha')^{3-p}} \over
       n \Omega_{n+1}}
    = {(2 \pi)^n \over n \Omega_{n+1}} \sqrt{\alpha'}^n g N_p \,,
  }
where $N_p$ is an integer and $\tau_p$ is the tension of a single
extremal D$p$-brane, and we have used the relation
  \eqn{NewtonG}{
   16 \pi G_N = 2 \kappa^2 = (2\pi)^7 g^2 (\alpha')^4 \,,
  }
valid in uncompactified type II~string theory.  The formula \QuantRA,
taken from \cite{hp}, can be checked against the standard formula for
D-brane tension:
  \eqn{DTension}{
   2 \kappa^2 \tau_p^2 = (2 \pi)^{7-2p} (\alpha')^{3-p} \,.
  }

Holding $N_p$ fixed, the behavior of the temperature in different
limits is well known.  In all cases, $T \to 0$ in the small $\alpha$
limit, since this is the large mass limit where the charge is
negligible.  For $p=1,2,3$, and $4$, $T \to 0$ also in the extremal
limit, whereas for $p=5$, $T$ approaches a constant, and for $p=6$,
$T$ diverges.  There are no other remarkable qualitative features of
the dependence of $T$ on mass other than a consequence of what we have
already said: for $p<5$, there is a special mass $M_*$ where $T$
reaches a global maximum.  This is the border of thermodynamic
stability: for masses less than $M_*$, the specific heat is positive,
and for masses greater than $M_*$ it is negative.  Holding $N_p$
fixed, one finds that \Tcl\ is extremized when
 \eqn{AlphaStr}{
  \sinh^2\alpha = \sinh^2\alpha_* = {1\over n-2} \,,
 }
or alternatively when
 \eqn{Mratio}{
  {M \over N_p \tau_p V} = {M_* \over N_p \tau_p V} = 
   {n^2-2 \over n\sqrt{n-1}}
 }
(note that $N_p \tau_p V$ is the extremal mass).  Another
characterization of the same condition is
 \eqn{rzeroStr}{
  r_0 = r_0^* = \left( {n-2 \over \sqrt{n-1}} \right)^{1/n} \tilde{R} \,,
 }
where $\tilde{R}$ is the length scale introduced in \QuantRA.

The M2 and M5-brane metrics are given by
 \eqn{MtwoMfiveMetric}{
  ds_{M2}^2 &= f_{M2}^{-2/3}\left( - h_{M2} dt^2
    + d\vec{x_2}^2 \right) + f_{M2}^{1/3}\left({dr^2 \over h_{M2}}
    + r^2 d\Omega_7^2 \right)  \cr
  ds_{M5}^2 &= f_{M5}^{-1/3}\left( - h_{M5} dt^2
    + d\vec{x_5}^2 \right) + f_{M5}^{2/3}\left({dr^2 \over h_{M5}}
    + r^2 d\Omega_4^2 \right) \,,
 }
where $f$ and $h$ are given by \HandF\ with $n=6$ for the M2-brane and
$n=3$ for the M5-brane.  Indeed, the thermodynamics for the M2-brane
and the D1-brane are identical.  This is not accidental: wrapping an
M2-brane around a circle gives a fundamental string of type IIA, which
is equivalent through some further dualities to the D1-brane (and,
more to the point, has the same Einstein metric up to a slightly
different identification of parameters).  For similar reasons, the
M5-brane and D4-brane have the same thermodynamics.  In particular,
the condition for the maximum Hawking temperatures for the M2-brane
and the M5-brane are given precisely by \AlphaStr, \Mratio, and
\rzeroStr, with $n=6$ and $3$, respectively.

\section{Universality classes and the general problem}
\label{universality}

A fairly broad statement of the question of $p$-brane stability in the
supergravity approximation can be summed up as follows: starting with
some $p$-brane solution in supergravity, one wishes to consider
perturbations that break some part of the translational invariance and
determine whether any of them grow with time.  Such considerations can
often be reduced to a problem in $2+1$ dimensions, and in the
following paragraphs we shall give more particulars about how this
comes about.

One starts with a two-derivative action in $\hat{D}>2$ spacetime
dimensions, containing Einstein gravity and perhaps also form fields
and scalars.  The brane in question is some solution with a horizon
and a weakly curved region far from it (typically asymptotically flat
space), where the whole solution admits as symmetries the Euclidean
group of translations and rotations in $p$
dimensions.\footnote{Smaller symmetry groups can arise.  In
``smeared'' brane solutions, for instance, composed of a density of
$p$-branes distributed in several orthogonal directions, the symmetry
group would be translations in all dimensions plus a product of two
rotation groups for the directions parallel to and perpendicular to
the individual $p$-branes.}  There should also be some Killing vector
which is timelike in the weakly curved region and null at the horizon,
which can be used to construct a notion of energy; and there may be
assorted spacelike Killing vectors as well, indicating rotational
symmetries.  The brane is of spatial co-dimension $\hat{D}-1-p$, and
far from the brane one generally expects an approximate
$SO(\hat{D}-1-p)$ rotational symmetry group.  The brane may have
various conserved charges and angular momenta (the latter
corresponding to generators of $SO(\hat{D}-1-p)$).  In this rather
general context, we would like to ascertain the stability properties
of the solution against small perturbations (determined by
thermodynamic stability, according to \cite{gmOne,gmTwo,Reall}),
perhaps also the dispersion relation for such perturbations, and
finally the end state of evolution along an unstable
direction---though this last is surely a much harder problem in
general than the other two.

Although we believe that the notion of universality classes we
introduce below will have some applicability to the general case
(including for example smeared brane solutions), let us focus on the
case where the $SO(\hat{D}-1-p)$ symmetry is exact.  Then
perturbations can be labeled uniquely by their wave-number $\vec{k}$
along the brane and their angular momentum quantum numbers under
$SO(\hat{D}-1-p)$.  It seems reasonable to assume that the $s$-wave
perturbations become unstable first, provided they exist as dynamical
perturbations (as opposed to being perturbations which are pure
gauge).  The perturbations in question thus preserve a large subgroup
of the symmetry of the original solution: $SO(\hat{D}-1-p)$ times that
part of the Euclidean group that preserves $\vec{k}$.  This makes it
particularly natural to consider a Kaluza-Klein ``reduction'' where
one integrates the action over an $S^{\hat{D}-2-p}$ orbit of
$SO(\hat{D}-1-p)$,\footnote{An interesting aside is that in
circumstances involving angular momentum, where obviously
$SO(\hat{D}-1-p)$ is not preserved even by the original solution,
Kaluza-Klein reduction is still often useful, with the somewhat
mysterious ``consistent truncation'' ansatze providing an exact
lower-dimensional account of both the solution and some low partial
wave perturbations \cite{gmOne,gmTwo}.  We suspect that the
coincidences arising in consistent truncation that matter for analyses
of horizon perturbations arise from bosonic symmetries rather than
supersymmetry.} and drops the integration over an ${\bf R}^{p-1}$
perpendicular to $\vec{k}$.  The result is a 2+1-dimensional action,
which in a fairly broad set of circumstances can be brought into the
form
 \eqn{Sgeneral}{
  S = \int d^3 x \, \sqrt{g} \left( R - {1 \over 2} G_{ij}(\vec{\phi})
   \partial_\mu \phi^i \partial^\mu \phi^j - V(\vec{\phi}) \right) \,.
 }
Einstein gravity, form fields, and scalars in $\hat{D}$ dimensions
reduce down to the type of action written in \Sgeneral, plus perhaps
abelian and non-abelian gauge fields.  The abelian gauge fields can be
dualized to scalars, unless there are Chern-Simons terms in the
action. Non-abelian gauge fields in general cannot be dualized, but
examples in the literature often have fields excited corresponding to
an abelian subgroup.  Thus \Sgeneral, while obviously not completely
general, does cover a broad range of cases.  Furthermore, it has been
observed \cite{gNaked} that $V(\vec{\phi})$ is commonly a sum of
exponentials of canonically normalized scalars: for example, the
integrated curvature of $S^{\hat{D}-2-p}$ scales as a power of its
radius, but the canonically normalized scalar measuring the size of
$S^{\hat{D}-2-p}$ is some multiple of the logarithm of the radius.

In section~\ref{GravityScalars} we will consider properties of
solutions to \Sgeneral, in part recapitulating arguments of
\cite{gNaked}.  In section~\ref{static} we briefly remark on the
reason to search for static perturbations.  Then in
section~\ref{CRsolns} we will make an explicit numerical study of
stability of a special case of the solutions considered in
section~\ref{GravityScalars}.

\subsection{Solutions to gravity plus scalars}
\label{GravityScalars}

Let us now study the slightly more general situation where 
 \eqn{SanyD}{
  S = \int d^D x \, \sqrt{g} \left( R - {1 \over 2} G_{ij}(\vec{\phi})
   \partial_\mu \phi^i \partial^\mu \phi^j - V(\vec{\phi}) \right) \,,
 }
where $D=d+1$ is unrelated to $\hat{D}$.  Motivated by the previous
discussion, we will be most interested in the case $D=3$.  Solutions
to the equations of motion of this action were studied in
\cite{gNaked}, of the form
 \eqn{SOoneone}{
  ds^2 = a(r)^2 (-dt^2 + d\vec{x}^2) + dr^2 \,,\qquad
  \vec{\phi} = \vec{\phi}(r) \,,
 }
with $a(r)$ a monotonically increasing function of $r$, running from
$0$ to $\infty$.  The large $a$ region in these solutions is weakly
curved (in fact, the main interest in \cite{gNaked} was asymptotically
anti-de Sitter space), and the small $a$ region is in most cases
singular.  These solutions have no horizons (or at best degenerate or
singular horizons), and their symmetry under boosts amounts to the
statement that they are extremal.  All such solutions can be derived
using a first order formalism (see for example
\cite{SkenderisTownsend,fgpwOne,dfgk}), which is a direct analog of
the Hamilton-Jacobi method of generating FRW cosmologies
\cite{SalopekBond}: if one starts with an action of the same form as
\Sgeneral, but in $D$ dimensions, and finds $W(\vec{\phi})$ such that
 \eqn{VW}{
  V(\phi) = {1\over 2} G^{ij} {\partial W \over \partial\phi^i}
   {\partial W \over \partial\phi^j} - {1\over 4} {D-1 \over D-2}W^2 \,,
 }
then a solution to
 \eqn{FirstOrder}{
  {\partial \phi^i \over \partial r}
    = - G^{ij} {\partial W \over \partial\phi^j} \,,
   \qquad {\partial a / \partial r \over a} 
    = {1\over 2(D-2)} W(\vec{\phi})
 }
will also solve the equations of motion.

Non-extremal generalizations of \SOoneone\ have the form
 \eqn{NonEx}{
  ds^2 = a(r)^2 (-h(r) dt^2 + d\vec{x}^2) + {dr^2 \over h(r)} \,,
   \qquad \vec{\phi} = \vec{\phi}(r) \,.
 }
We assume in general that $h(r) \to 1$ in the region where $a(r) \to
\infty$, and $h(r) < 1$ everywhere.  If $h(r_H)=0$ and $r_H$ is the
largest value of $r$ for which this is true, then $r=r_H$ is an event
horizon with respect to the region where $a(r) \to \infty$ (which is
also at large $r$).  Provided $h'(r_H) \neq 0$, this is a
non-degenerate horizon with finite Hawking temperature.

Solutions of the form \NonEx\ involve a number of constants of
integration.  The counting of them has been explained in
\cite{gNaked}, and we will recapitulate briefly here.  The relevant
equations of motion are all ordinary differential equations in $r$
once we assume the form \NonEx.  They comprise the second order
equations for the scalars, of which we suppose there are $n$; the
$G_{rr}$ component of Einstein's equation, which is a first order
equation; and one further Einstein equation (for instance, the
combination $G^t{}_t - G^x{}_x$ of Einstein's equations), which is
second order.  There are $2n+3$ integration constants altogether.  One
amounts to an additive shift in the radial variable $r$, which is just
a coordinate transformation.  Another is fixed by the requirement that
$h(r) \to 1$ in the region where $a(r)$ is large.  The presence of a
non-degenerate horizon fixes $n$ more, as explained in \cite{gNaked}
(one way to understand the presence of these $n$ horizon constraints
is that in the Wick-rotated Euclidean solution, where the horizon
becomes a point, the scalars must be smooth everywhere, so their
radial derivatives must vanish at this point).  This leaves $n+1$
parameters that specify the solution.  One of these is the temperature
of the horizon, and the other $n$ pertain to the asymptotics of the
scalars in the large $a$ region.  Of these $n$ parameters, some but
not all may be fixed by demanding regularity in the large $a$ region.
In an $AdS_{d+1}/CFT_d$ context, where the large $a$ region is
asymptotically anti de-Sitter, the $n$ parameters correspond to the
coefficients of $e^{-(d-\Delta_i) r/L}$ in the expansion of the
scalars around their constant limiting values as $r \to \infty$.
These coefficients amount on the CFT side to the mass parameters of
gauge singlet operators added to the lagrangian.  If some of the
operators in question are irrelevant, then the corresponding scalars
have positive $m^2$, and the larger of the two solutions to the
linearized scalar equations of motion blows up at large $r$: this is a
case where parameters are fixed to zero by demanding regularity in the
large $a$ region.  In the case where the dual operators are relevant,
and the corresponding scalars have negative $m^2$, then the
coefficient of $e^{-(d-\Delta_i) r/L}$ is truly a variable parameter.
In the case of asymptotically anti-de Sitter solutions, an additional
scaling symmetry can be used to eliminate one additional parameter:
this amounts to applying a scale transformation on the CFT side to
set, say, the temperature equal to $1$.

Let us now consider the situation where the asymptotics of the scalars
is entirely fixed in the large $a$ region, leaving us with one free
parameter.  One can show (see \cite{gNaked} for the case where $d=4$)
that
 \eqn{hSolve}{
  h = 1 - B \int_r^\infty {dr_1 \over a(r_1)^d} \,,
 }
and then $B$ is the free parameter.  (The identity \hSolve\ holds even
though $a(r)$ may in general change when $B$ changes.  It is a
convenient parametrization because the influence of $B$ on $a(r)$ is
small in the large $a$ region).  In an $AdS_{d+1}/CFT_d$ context, this
would correspond to specifying a deformation of the gauge theory
lagrangian by relevant operators and then varying either the
temperature or the energy density.  Evidently, in the limit $B \to 0$,
one recovers a solution of the form \SOoneone.  Suppose that in this
limiting solution, $V(\vec{\phi}(r))$ decreases monotonically as
$a(r)$ decreases, going to $\infty$ as $a \to 0$.\footnote{It is
possible that this condition could be relaxed somewhat: we must
certainly have $V(\vec{\phi}(r))$ less than its limiting value at
large $a$, and probably the monotonicity property is also needed for
$a$ less than a certain upper bound.}  It can then be argued (though
not wholly rigorously) \cite{gNaked} that
 \begin{itemize}
  \item Solutions with regular horizons exist for all $B>0$.
  \item If $V(\vec{\phi})$ is the sum of exponentials of canonically
normalized scalars, then in the $B=0$ solution, the scalars have an
asymptotic direction as $a(r) \to 0$: that is, $\vec{\phi}(r) \sim
\vec{\phi}_0 + \vec{v} \varphi(r)$ for some constants $\vec{\phi}_0$
and $\vec{v}$, which we may normalize so that $\varphi$ is itself
canonically normalized.  Then $V(\vec{\phi}(r)) \approx
e^{2\gamma\varphi}$ in the region of small $a$.
  \item In near-extremal solutions, with sufficiently small $B$, the
solution at small $a$ is well-approximated by a Chamblin-Reall
solution for the exponent $\gamma$, which we explain below in
section~\ref{CRsolns}.  The whole solution can be well-approximated by
the original extremal solution patched onto a Chamblin-Reall solution.
These approximations become progressively better as $B \to 0$.
  \item The entropy and temperature of near-extremal solutions are
related by a power law: $S \propto V T^\alpha$, where 
   \eqn{alphagamma}{
    \alpha = {D-2 \over 1 - 2 \gamma^2(D-2)} \,.
   }
 \end{itemize}

These features suggest the notion of universality class that we are
going to use.  Suppose that an extremal solution to an action with
canonically normalized scalars has a ``scaling region'' where
$\vec{\phi}(r) \approx \vec{\phi}_0 + \vec{v} \varphi(r)$ and $V
\approx e^{2\gamma\varphi}$ for many decades of variation of $a(r)$.
Then non-extremal solutions with the horizon well within the scaling
region should again be well-approximated by the original extremal
solution patched onto a Chamblin-Reall solution near the horizon.  The
approximation should become good in the limit where the scaling region
in the extremal solution extends far above and below the value of $a$
where the horizon is located in the non-extremal solution.  The same
power law, $S \propto V T^\alpha$, should pertain.

With the idea of universality classes in hand, we can address
stability of the horizon in a simple way.  In section~\ref{CRsolns}, we
will study explicitly perturbations around Chamblin-Reall solutions,
involving the metric and the ``active'' scalar $\varphi$.  We will
find perturbations which are normalizable both at the horizon and far
from it.  As the full solution is well approximated by matching the
extremal solution onto a Chamblin-Reall solution, perturbations
should be well approximated by similarly matched solutions---only,
because the perturbations we are considering are normalizable within
the Chamblin-Reall region, they approach zero near the matching
region, and the perturbations of fields in the extremal region are
very small.  It does not seem plausible that there are normalizable
perturbations which are large in the extremal region and small as one
enters the Chamblin-Reall region near the horizon.  To sum up, we
believe the stability properties of the whole solution are determined
by the stability properties of the near-horizon Chamblin-Reall
solution, which also determines the scaling of entropy with
temperature.

It may seem that we are focusing rather narrowly on a rather special
class of perturbations: not only invariant under the $SO(\hat{D}-1-p)$
that we started with in carrying out the Kaluza-Klein reduction, but
also involving only the metric and the active scalar near the horizon.
This amounts to focusing on a type of fluctuations which we might
describe as adiabatic, since the fields are varying locally in the
same proportions that they would do globally if we simply changed the
temperature.  But this is precisely the mode of instability that the
thermodynamic arguments of \cite{gmOne,gmTwo,Reall} suggest.
Instabilities in other modes are conceivable, and one could even study
them by considering scalar fluctuations ``transverse'' to the
solution.  But we would find it very surprising if such fluctuations
gave rise to normalizable instabilities when the fluctuations we
consider do not.  To put it another way, for that to happen, the logic
of \cite{Reall} must fail.  Our rather detailed predications about the
nature of the instability amount in a sense to an elaboration of the
connection between thermodynamic and dynamical instabilities.

\subsection{Static perturbations}
\label{static}

In the rest of this paper, we will investigate Gregory-Laflamme type
instabilities of various black branes.  For all the unstable black
branes Gregory and Laflamme analyzed \cite{glOne,glTwo}, they found at
linear order in perturbation theory that there is a static
perturbation, and that instabilities occur at longer wavelengths than
this static perturbation.

As remarked in the introduction, Reall \cite{Reall} in his approach to
the proof of the conjecture of \cite{gmOne,gmTwo} investigated the
role of such time-independent perturbations, and argued that they are
of central importance to the relation between thermodynamic and
dynamical instability. In particular he related such perturbations of
a black-brane solution to the negative modes of the Lichnerowicz
operator on the Euclidean black hole background found by compactifying
the black $p$-brane on a $p$-torus and doing a Wick rotation. The
existence of such negative modes is then related to thermodynamic
stability. (Negative modes of Euclidean black holes have also been
analyzed in the context of finite temperature stability of
flat spacetime by Gross, Perry and Yaffe \cite{gpy}.)

Thus when looking for instabilities of black branes, instead of
considering instabilities with arbitrary growth rates, we will
restrict our attention to static threshold perturbations.  With the
relation between thermodynamic and dynamical stabilities in mind, we
expect the dispersion curve of a Gregory-Laflamme type instability to
end at such a threshold point at the small wavelength limit.

\subsection{The Chamblin-Reall solutions}
\label{CRsolns}

As argued in section~\ref{GravityScalars}, close to the horizon, black
brane solutions \cite{cr} for gravity coupled to a scalar $\phi$ with
a potential of the form $V=V_0 e^{2\gamma\phi}$ provide good
approximations to a broad range of supergravity solutions, including
ones that are non-extremal versions of those that represent RG-flows
in AdS/CFT.

The action for the gravity plus scalar system is
  \eqn{GravScl}{
    S &= \int d^{D}x \sqrt{g} \left(R  - {1\over 2}(\partial \phi)^2 -
        V(\phi) \right) \cr
    V(\phi) &= V_0 e^{2\gamma\phi} \,.
  }
We will restrict our attention to negative $V_0$, since that is the
case that admits black hole solutions. The black brane solutions are given 
by
  \eqn{crAnsatz}{
    ds_0^2 = a_0^2(r)(-h_0(r) dt^2 + d\vec{x}_{D-2}^2) + {dr^2\over h_0(r)} \,,
  }
where
  \eqn{crSoln}{
    h_0(r) &= 1 - 
     \left( {r_0\over r} \right)^{{D - 1\over 2 \gamma^2 (D - 2)} - 1} \cr
    a_0(r) &= r^{1\over 2 \gamma^2(D - 2)} \cr
    \phi_0(r) &=  -{1\over \gamma} \log(\gamma^2 c r) \,,
  }
and $c$ is given by the positive root of the equation
  \eqn{cgammaV}{
    V_0 = c^2 \left( {1\over 2} \gamma^2 - {1\over 4} {D-1\over D-2} \right) \,.
  }
In the notation of \cite{cr} (differing slightly from ours), these
solutions correspond to the ``type II'' case, with $b^2 < D-1$, $M>0$,
and $V_0 < 0$.  Note that in the gauge presented above, the metric is
independent of $V_0$: it only depends on $\gamma$, whereas the scalar
profile depends on both $\gamma$ and $V_0$.

With respect to the general discussion of non-extremal horizons in
section~\ref{GravityScalars}, the solution \crSoln\ is rather special,
in that the functions $a(r)$ and $\phi(r)$ are entirely independent of
the parameter $r_0$ that determines the location of the horizon.  (The
equivalent parameter in section~\ref{GravityScalars} is $B$).  In
other words, one can get at the solution \crSoln\ by first obtaining
the extremal solution and then using \hSolve\ without changing $a(r)$.
In general, $h \neq 1$ ``back-reacts'' on the solution, changing
$a(r)$ and $\phi(r)$. This property of \crSoln\ also hints at
the origin of the relation \cgammaV : One plugs $W = ce^{\gamma\phi}$ in 
\VW\ to obtain $V$, just as one would do when looking for 
extremal solutions.

As noted in \cite{cr},
for a discrete set of $\gamma$'s it is possible to
arrive at the solutions \crSoln, \cgammaV\ by a dimensional reduction of 
spacetimes of the form
$Schwarzchild_{q+2} \times R^{D-2}$ on the $q$-sphere of
$Schwarzchild_{q+2}$. For such cases, $V_0$ and $\gamma$ are given in
terms of $q$ as
  \eqn{V0gamma}{
    V_0 &= -q(q-1) \cr
    \gamma &= \sqrt{D+q-2 \over 2q(D-2)} \,.
  }
At least for this discrete set of solutions, we certainly expect to
find dynamical instabilities: they correspond to the instabilities of
an uncharged black string in $q+3$ spacetime dimensions.  The
$\gamma$'s that can be obtained by a dimensional reduction on a
$q$-sphere, where $q$ runs from $1$ to $\infty$, fall in the range
  \eqn{gammaRange}{
    \sqrt{1 \over 2(D-2)} < \gamma < \sqrt{D-1 \over 2(D-2)} \,.
  }
(The case $q=1$, which gives the upper limit on $\gamma$, is
degenerate: $V_0 = 0$ for that case.) It turns out that \gammaRange\
is precisely the range for which the $D$-dimensional black brane has a
negative specific heat. The conjecture of \cite{gmOne,gmTwo} then
implies that one should have dynamical instabilities for the $\gamma$'s
precisely in this range, including the ones that {\it can't} be
obtained by dimensional reduction.  In the next section, we will give
numerical evidence in favor of this for the case $D=3$.  (The same
conclusion for other $D$ then follows from straightforward
Kaluza-Klein reduction).

\subsection{Numerical study of the stability of Chamblin-Reall
solutions}

Choosing $D=3$, we searched numerically for threshold dynamical
instabilities of the Chamblin-Reall solutions by restricting attention
to perturbations of the form
  \eqn{PertCR}{
    ds^2 &= -e^{2A(r,x)} \left(a_0^2(r) h_0(r) dt^2\right) + e^{2B(r,x)}
      \left(a_0^2(r) dx^2 + {dr^2 \over h_0(r)}\right) \cr
    \phi(r) &= \phi_0(r) + \lambda \cos{kx} \phi_1(r) \,,
  }
where
  \eqn{CRab}{
    A(r,x) &= \lambda \cos{kx} A_1(r) \cr
    B(r,x) &= \lambda \cos{kx} B_1(r) \,,
  }
and working to linear order in $\lambda$. It is possible to show that this
ansatz for the perturbations is consistent with the equations of motion.

The Einstein equations in $D=3$ read
  \eqn{EE}{
    R^\mu_\nu = T^\mu_\nu - g^\mu_\nu T^\alpha_\alpha \,.
  }
Four of these equations are nontrivial, namely the ones involving
$R^t_t$, $R^r_r$, $R^x_x$ and $R^r_x$.  Using the $R^r_x$ equation,
one can solve algebraically for $B_1$ in terms of $A_1$, $A_1'$, and
$\phi_1$.  The $R^t_t$ equation and the scalar equation of motion
involve $B_1$, but not its derivatives. Plugging in $B_1$ as found
from the $R^r_x$ equation into the $R^t_t$ equation and the scalar
equation of motion, we end up with the following two equations
involving $A_1$ and $\phi_1$:
  \eqn{CReoms}{
    0 &= \left( -\left( {\gamma}^6\,{k}^2\,r^2 \right)  + r^{{\gamma}^{-2}} - 2\,{\gamma}^2\,r^{{\gamma}^{-2}} + 
     {\gamma}^4\,r^{{\gamma}^{-2}}\,\left( 1 + {k}^2\,r \right)  \right) \,A_1(r)
 + 
  \gamma\,{\left( -1 + {\gamma}^2 \right) }^2\,r^{{\gamma}^{-2}}\,\phi_1(r) \cr 
& \qquad{}+ 
  \frac{1}{2}\gamma^2\,\left( -3\,{\gamma}^4\,r^2 + r^{{\gamma}^{-2}}\,\left( -4\,r + r^{{\gamma}^{-2}} \right)  - 
       2\,{\gamma}^2\,r^{{\gamma}^{-2}}\,\left( -5\,r + 2\,r^{{\gamma}^{-2}} \right)  \right) \,A_1'(r) \cr
& \qquad{} - 
  {\gamma}^4\,r\,\left( r - r^{{\gamma}^{-2}} \right) \,\left( {\gamma}^2\,r - r^{{\gamma}^{-2}} \right) \,
   A_1''(r) \cr
    0 &= 2\,{\left( -1 + {\gamma}^2 \right) }^2\,r^{{\gamma}^{-2}}\,A_1(r) + 
  \gamma\,\left( 2\,r^{{\gamma}^{-2}} - {\gamma}^2\,r^{{\gamma}^{-2}}\,\left( 4 + {k}^2\,r \right)  + 
     {\gamma}^4\,\left( {k}^2\,r^2 + 2\,r^{{\gamma}^{-2}} \right)  \right) \,\phi_1(r) \cr
& \qquad{} - 
  {\gamma}^2\,\left( -r + r^{{\gamma}^{-2}} \right) \,
   \left( -3\,r^{{\gamma}^{-2}} + {\gamma}^2\,\left( -r + 4\,r^{{\gamma}^{-2}} \right)  \right) \,A_1'(r) + 
  \gamma\,{\left( -\left( {\gamma}^2\,r \right)  + r^{{\gamma}^{-2}} \right) }^2\,\phi_1'(r) \cr
& \qquad{} + 
  {\gamma}^3\,r\,\left( -r + r^{{\gamma}^{-2}} \right) \,\left( -\left( {\gamma}^2\,r \right)  + r^{{\gamma}^{-2}} \right) \,
   \phi_1''(r) \,.
  }
(It is possible to do a consistency check by arriving at the same
equations using the $R^x_x$ and $R^r_r$ equations).

\begin{figure}
  \centerline{\psfig{figure=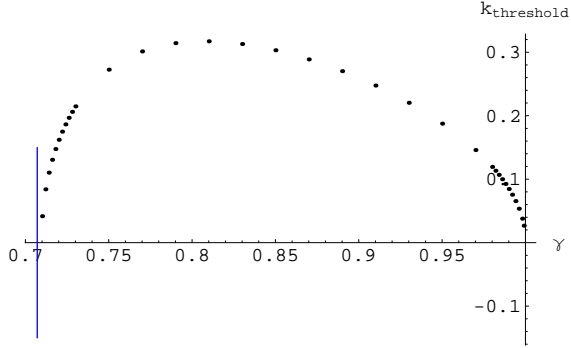,width=3in}}
  \caption{Stability of Chamblin-Reall solutions. The dots represent 
     threshold instabilities of the Chamblin-Reall solutions. The vertical 
     line at $\gamma^*=1/\sqrt{2}$ indicates the thermodynamic 
     stability limit. 
     }\label{figCR}
 \end{figure}

We used the linearity of the equations to fix the value of $\phi_1$ at
the horizon.  Then, using a a shooting algorithm, we looked for values
of $k$ and $A_1(r_0)$ that would give solutions that are well-behaved
at the horizon and infinity, for a given $\gamma$.  The results for
$k$ as a function of $\gamma$ are shown in figure~\ref{figCR}. As can
be seen from the figure, the wavelengths of the threshold
instabilities diverge as one approaches the thermodynamic stability
limit given by $\gamma=\gamma^* \equiv {1\over \sqrt{2}}$ (the lower
limit in \gammaRange).  $\gamma=1$, which corresponds to the upper
limit in \gammaRange, is where the causal structure of the
Chamblin-Reall solutions changes: in the notation of Chamblin and
Reall \cite{cr}, going from $\gamma<1$ to $\gamma>1$ corresponds to
going from the $b^2<D-1$ case to the $b^2>D-1$ case of the type~II
solutions.

For $\gamma < 1/\sqrt{2}$, we found no static perturbation.  This
plus the thermodynamic stability of the solutions strongly suggests
that they are stable.

In order to address the question of what happens near the threshold of
instability, we attempted to fit the behavior of $k$ as a function of
$\gamma$ to a power law near $\gamma=\gamma^*$.  The fit was
considerably better if the threshold of stability was shifted to
$\gamma^*_{num} \approx 1.003 \gamma^*$: then we found $k \sim
\sqrt{\gamma-\gamma^*_{num}}$.  The fit to a power law was still less
than spectacular.  We suspect that numerical error contributed both to
this and to the discrepancy between $\gamma^*$ and $\gamma^*_{num}$.

\section{Crossing the threshold of stability: non-dilatonic\\ branes}
\label{Cross}

Next, we would like to investigate the dynamical stability of D3, M2
and M5-branes. Following the philosophy introduced in the beginning of
section~\ref{universality}, we will do a Kaluza-Klein reduction of the
metrics \SMetric\ and \MtwoMfiveMetric\ on all but one of the spatial
worldvolume directions and $S^{n+1}$, to end up with a 3 dimensional
background.  We will be working with $s$-wave perturbations of this
background, keeping only the two scalars that describe the sizes of
$S^{n+1}$ and the torus, ignoring all the higher harmonics. There will
be two contributions to the potential of these scalars: one due to the
curvature of the $S^{n+1}$ that we compactify on, another due to the
kinetic term of the 4-form (3-form) potential in IIB
(eleven-dimensional) supergravity. Both of these terms will turn out
to be exponentials of a linear combination of the two scalars, hence
the notion of universality classes introduced in
section~\ref{universality} will be relevant in this context.

\subsection{Kaluza-Klein reduction}

We want to perform the dimensional reduction in a way that would give
an Einstein-Hilbert action for gravity, without a scalar-dependent
coefficient in front of the three-dimensional Ricci scalar. For this
purpose, let's start with a general metric
 \eqn{GenMet}{
   ds_D^2 = d\bar{s}_r^2 + e^{2K} ds_p^2 + e^{2L}ds_q^2
 }
where $D = p + q + r$, $d\bar{s}_r^2$ is a metric on the Lorentzian $r$-manifold
$X_r$, $ds_p^2$ and $ds_q^2$ are metrics on compact Riemannian manifolds
$Y_p$ and $Z_q$, respectively, and $K$ and $L$ are functions that depend
only on the coordinates of $X_r$. For example, for the case of the D3-brane,
$p=2$, $Y_p = T^2$ (two worldvolume directions are compactified), $q=5$,
$Z_q = S^5$, $r=3$. Defining
 \eqn{RescMet}{
   d\bar{s}_r^2 = e^{2N} ds_r^2 \,,
 }
with
 \eqn{DefN}{
   N = - {pK +qL \over r - 2} \,,
 }
one gets
 \eqn{GenDimRed}{
   \sqrt{-g_D}&R(ds_D^2) = \sqrt{-g_r}\sqrt{g_p}\sqrt{g_q}
     \Big[R(ds_r^2) + e^{2(N-K)} R(ds_p^2) + e^{2(N-L)} R(ds_q^2) - \cr
        &{1 \over r-2}\big(
          p(r+p-2) \partial_\mu K \partial_\nu K g_r^{\mu\nu} +
          q(r+q-2) \partial_\mu L \partial_\nu L g_r^{\mu\nu} +
          2pq      \partial_\mu K \partial_\nu L g_r^{\mu\nu}
          \big)
     \Big] \,,
 }
where $R(ds^2)$ denotes the Ricci scalar for the metric $ds^2$. The first term
gives the Einstein-Hilbert lagrangian for $ds_r^2$ (without a scalar-dependent
factor---as we wanted), and the rest are the kinetic and potential terms for the
scalars $K$ and $L$. In our applications, $R(ds_p^2)$ and $R(ds_q^2)$
will be the 
Ricci scalars of $T^p$ and $S^q$, respectively, i.e.~0 and $q(q-1)$.

Using these results, we get the dimensionally reduced
form of the D3-brane metric and the scalars as
  \eqn{RedDThreeSol}{
    ds_0^2 &= - r^{10}f h dt^2 + r^{10} {f^2 \over h} dr^2 + r^{10} f dx^2 \cr
    K_0(r) &= -{1 \over 4} \log{f} \cr
    L_0(r) &= \log{r} + {1\over 4} \log{f}
  }
where $f$ and $h$ are given by \HandF\ with $n=4$. The three-dimensional
action is given by
  \eqn{RedDThreeAct}{
    S = \int d^3x \sqrt{g_3} \left(R - {1\over 2} \partial_\mu \phi^i
      \partial_\nu \phi^j G_{ij} g_3^{\mu\nu} - V(\vec{\phi}) \right)
  }
where
  \eqn{DThreeKinMat}{
    G_{ij} =
      \left( \begin{array}{cc}
      12 & 20  \\
      20 & 60  \end{array} \right)  \qquad\qquad
    \vec{\phi} =
      \left( \begin{array}{c}
      K \\ L
      \end{array} \right)
  }
  \eqn{DThreePot}{
    V(\vec{\phi}) = -20 e^{-4K-12L} + 8\tilde{R}_{D3}^8 e^{-4K-20L} \,,
  }
and $\tilde{R}_{D3}$ is the length scale defined in \QuantRA, with
$n=4$.  The first term in $V$ comes from \GenDimRed, and the second
one comes from the dimensional reduction of the $F_5^2$ term in the
IIB SUGRA action.  We have omitted overall factors from \RedDThreeAct,
and we have introduced notation that somewhat obscures dimensional
checks: for instance, $e^L$ has the dimensions of length.  In
practice, we will simply choose an arbitrary numerical value for
$\tilde{R}_{D3}$ for the numerics in section~\ref{CrossNumerics}.
Another way to think of this is that we choose an arbitrary value for
the Planck length, which is permissible because our analysis is
entirely classical.

Since the second term in the potential is due to the charge of the
D3-brane, we expect this term to be of little significance in the
large non-extremality limit.  By performing a linear transformation of
the scalars $K$ and $L$ to canonically normalized ones (i.e. those
with $G_{ij} = \delta_{ij}$ in equation \SanyD) one of which is aligned
with the exponent in the first term of the potential, one finds that
this term corresponds to a Chamblin-Reall coefficient
$\gamma=\sqrt{3\over 5}$. That this value is in the unstable range
\gammaRange\ is consistent with expectations, since the brane is
expected to be unstable when the effects of the charge are negligible.

Another consistency check can be done by calculating the relevant
exponent $\alpha$ in the Chamblin-Reall equation of state by using
\alphagamma. The equation of state of an uncharged brane can be read
off from \BekHawk\ and \Tcl\ to be $S\propto VT^{-(n+1)}$. The
exponent obtained from \alphagamma\ by setting $\gamma=\sqrt{3\over 5}$
agrees precisely with this: one gets $S\propto VT^{-5}$ in
both cases.

When the brane is near-extremal, $R^4\equiv r_0^4\sinh^2{\alpha}
\gg r_0^4$, so for radii close to the horizon, $R\gg r$. Using this in
\RedDThreeSol, we see that $L_0$ is approximately constant near the
horizon, i.e. for a near-extremal D3-brane the ``active'' scalar close
to the horizon is $K$. Thus, in order to read off the relevant
Chamblin-Reall exponent $\gamma$ in the near-extremal regime, one sets
$L=const$ in the action, and rescales $K$ to make it canonically
normalized: $\tilde{K}=\sqrt{12}K$. Then, both terms in the potential
become a multiple of $e^{-{2 \tilde{K}\over\sqrt{3}}}$ which shows
that $\gamma={1\over\sqrt{3}}$.\footnote{The sign of $\gamma$ is
arbitrary since the field redefinition $\phi\to-\phi$ is equivalent to
switching the sign of $\gamma$ in \GravScl, \crSoln\ and
\cgammaV.}  This value of $\gamma$ is in the thermodynamically stable
range, as expected for a near extremal D3-brane. Equation \alphagamma\
gives the equation of state of the near-extremal D3-brane to be
$S\propto VT^{3}$, which is the equation of state of a 3+1-dimensional
CFT, in line with the AdS/CFT correspondence.

The expected transition from the stability for near-extremal branes to
instability at large non-extremalities is also confirmed by the
numerical analysis of dynamical instabilities, which we present in the
next section.

The analysis for black M2 and M5-branes is similar. M2-brane versions of   
\RedDThreeSol, \DThreeKinMat, and \DThreePot\ are 
  \eqn{MTwoEqns}{
    ds_0^2 &= - r^{14}f h dt^2 + r^{14} {f^2 \over h} dr^2 + r^{14} f dx^2 \cr
    K_0(r) &= -{1 \over 3} \log{f} \cr
    L_0(r) &= \log{r} + {1\over 6} \log{f}\cr
    G_{ij} &= 
      \left( \begin{array}{cc}
       4 &  14  \\
      14 & 112  \end{array} \right) \cr
    V(\vec{\phi}) &= -42 e^{-2K-16L} + 18 
     \tilde{R}_{M2}^{12} e^{-2K-28L} \,, \cr
  }
and M5-brane versions are
  \eqn{MFiveEqns}{
    ds_0^2 &= - r^{8}f h dt^2 + r^{8} {f^2 \over h} dr^2 + r^{8} f dx^2 \cr
    K_0(r) &= -{1 \over 6} \log{f} \cr
    L_0(r) &= \log{r} + {1\over 3} \log{f}\cr
    G_{ij} &=
      \left( \begin{array}{cc}
      40 & 32  \\
      32 & 40  \end{array} \right) \cr
    V(\vec{\phi}) &= -12 e^{-8K-10L} + {9\over 2} 
     \tilde{R}_{M5}^6 e^{-8K-16L} \,,
  }
where $h$ and $f$ are given by \HandF\, and $\tilde{R}_{M2}$ and
$\tilde{R}_{M5}$ are given by \QuantRA\ with $n=6$ for the M2-brane
and $n=3$ for the M2-brane. Once again for large non-extremality the
potential terms due to the charges can be neglected, and by using
canonically normalized scalars one can calculate the $\gamma$'s that
are relevant for the terms due to the curvatures of $S^{n+1}$. One
gets $\gamma={2\over\sqrt{7}}$ for the M2-brane and
$\gamma={\sqrt{5}\over2\sqrt{2}}$ for the M5-brane. As in the D3-brane
case, these are consistent with the equations of state of uncharged
branes: $S\propto VT^{-7}$ for the M2-brane and $S\propto VT^{-4}$ for
the M5-brane. In the near-extremal limits, the ``active'' scalar is
$K$ for the M2 and M5-branes as well, and one gets the corresponding
Chamblin-Reall coefficients as $\gamma={1\over 2}$ for the M2-brane
and $\sqrt{2\over 5}$ for the M5-brane. The equations of state one
gets from these are consistent with expectations from AdS/CFT:
$S\propto VT^2$ for the M2-brane and $S\propto VT^5$ for the M5-brane.

\subsection{Numerical stability analysis for non-dilatonic branes}
\label{CrossNumerics}

We will describe the stability analysis for the D3-brane in some
detail, and just show results for the M2-brane and the M5-brane.

In order to investigate the dynamical stability of the
three-dimensional background \RedDThreeSol, obtained by Kaluza-Klein
reduction from the D3-brane, we introduce the perturbations
  \eqn{DThreePert}{
    ds^2 &= - e^{2A(r,x)} \left( r^{10} f h dt^2 \right) + e^{2B(r,x)} \left(
      r^{10}{f^2 \over h} dr^2 + r^{10} f dx^2 \right) \cr
    A(r,x) &= \lambda \cos{kx} A_1(r) \cr
    B(r,x) &= \lambda \cos{kx} B_1(r) \cr
    K(r,x) &= K_0(r) + \lambda \cos{kx} K_1(r) \cr
    L(r,x) &= L_0(r) + \lambda \cos{kx} L_1(r) \,,
  }
and work to linear order in $\lambda$.  $K$ and $L$ are the scalars
introduced in \RedDThreeAct.

There are four nontrivial Einstein equations (those involving $R^t_t$,
$R^r_r$, $R^x_x$ and $R^r_x$), and two scalar equations of
motion. Using the $R^r_x$ equation, one can solve for $B_1 = B_1(A_1,
A_1^\prime, K_1, L_1)$.  The $R^t_t$ equation and both of the scalar
equations of motion involve $B_1$, but not its derivatives.  Plugging
in $B_1$ as found from the $R^r_x$ equation into the $R^t_t$ equation
and the scalar equations of motion, we end up with three equations
involving $A_1$, $K_1$ and $L_1$ (it is possible to do a consistency
check by arriving at the same equations using the $R^x_x$ and $R^r_r$
equations). Defining $R^4 = r_0^4 \sinh^2{\alpha}$, the equations read

\eqn{EOMOne}{
0 =& r\,\Bigg( 16\,r^2\,{r_0}^4\,\left( 5\,r^8 + 10\,r^4\,R^4 + 3\,R^8 - 2\,R^4\,{r_0}^4 \right)  \cr
& \qquad{} + 
     {k}^2\,{\left( r^4 + R^4 \right) }^2\,\left( 5\,r^8 - R^4\,{r_0}^4 + 3\,r^4\,\left( R^4 - {r_0}^4 \right)  \right) 
     \Bigg) \,A_1(r) \cr
& - 32\,r^3\,{r_0}^4\,
   \left( 5\,r^8 + 10\,r^4\,R^4 + 3\,R^8 - 2\,R^4\,{r_0}^4 \right) \,K_1(r) \cr
& - 
  160\,r^3\,\left( 2\,r^4\,R^4\,{r_0}^4 + R^8\,{r_0}^4 + 
     r^8\,\left( 2\,R^4 + 3\,{r_0}^4 \right)  \right) \,L_1(r) \cr
& - 
  2\,\Bigg( 5\,r^{16} + R^8\,{r_0}^8 - 5\,r^{12}\,\left( R^4 + {r_0}^4 \right)  + 
     r^8\,\left( 20\,R^4\,{r_0}^4 + 6\,{r_0}^8 \right)  \cr
& \qquad{} + 
     r^4\,\left( 5\,R^8\,{r_0}^4 - 3\,R^4\,{r_0}^8 \right)  \Bigg) \,A_1'(r) \cr
& - 
  r\,{\left( r^4 + R^4 \right) }\,\left( r^4 - {r_0}^4 \right) \,
   \left( 5\,r^8 - R^4\,{r_0}^4 + 3\,r^4\,\left( R^4 - {r_0}^4 \right)  \right) \,A_1''(r)
}

\eqn{EOMTwo}{
0 =& 8\,r^3\,{r_0}^4\,\left( 10\,{\left( r^4 + R^4 \right) }^2 - 4\,R^4\,\left( R^4 + {r_0}^4 \right)  \right) \,
   A_1(r) \cr
& + \Bigg( -32\,r^3\,{r_0}^4\,\left( 5\,r^8 + 10\,r^4\,R^4 + 3\,R^8 - 2\,R^4\,{r_0}^4 \right) \cr
& \qquad{} + 
     3\,{k}^2\,r\,{\left( r^4 + R^4 \right) }^2\,\left( 5\,r^8 - R^4\,{r_0}^4 + 3\,r^4\,\left( R^4 - {r_0}^4 \right)  \right) 
     \Bigg) \,K_1(r) \cr
& + 5\,\Bigg( {k}^2\,r\,{\left( r^4 + R^4 \right) }^2\,
      \left( 5\,r^8 - R^4\,{r_0}^4 + 3\,r^4\,\left( R^4 - {r_0}^4 \right)  \right)  \cr
& \qquad{} - 
     32\,r^3\,\left( 2\,r^4\,R^4\,{r_0}^4 + R^8\,{r_0}^4 + r^8\,\left( 2\,R^4 + 3\,{r_0}^4 \right)  \right)  \Bigg) 
    \,L_1(r) \cr
& + \left( r^4 - {r_0}^4 \right) \,
   \left( 15\,r^{12} + 3\,R^8\,{r_0}^4 + 5\,r^8\,\left( 10\,R^4 + 3\,{r_0}^4 \right)  + 
     r^4\,\left( 15\,R^8 - 2\,R^4\,{r_0}^4 \right)  \right) \,A_1'(r) \cr
& - 
  3\,\left( r^4 + R^4 \right) \,\left( 5\,r^4 - {r_0}^4 \right) \,
   \left( 5\,r^8 - R^4\,{r_0}^4 + 3\,r^4\,\left( R^4 - {r_0}^4 \right)  \right) \,K_1'(r) \cr
& - 
  5\,\left( r^4 + R^4 \right) \,\left( 5\,r^4 - {r_0}^4 \right) \,
   \left( 5\,r^8 - R^4\,{r_0}^4 + 3\,r^4\,\left( R^4 - {r_0}^4 \right)  \right) \,L_1'(r) \cr
& - 
  3\,\left( r^4 + R^4 \right) \,\left( r^5 - r\,{r_0}^4 \right) \,
   \left( 5\,r^8 - R^4\,{r_0}^4 + 3\,r^4\,\left( R^4 - {r_0}^4 \right)  \right) \,K_1''(r) \cr
& - 
  5\,\left( r^4 + R^4 \right) \,\left( r^5 - r\,{r_0}^4 \right) \,
   \left( 5\,r^8 - R^4\,{r_0}^4 + 3\,r^4\,\left( R^4 - {r_0}^4 \right)  \right) \,L_1''(r)
}

\eqn{EOMThree}{
0 = & 8\,r^3\,{r_0}^4\,\left( 6\,{\left( r^4 + R^4 \right) }^2 - 4\,R^4\,\left( R^4 + {r_0}^4 \right)  \right) \,
   A_1(r) \cr
& + \Bigg( -32\,r^3\,{r_0}^4\,\left( 3\,r^8 + 6\,r^4\,R^4 + R^8 - 2\,R^4\,{r_0}^4 \right) \cr
& \qquad{} + 
     {k}^2\,r\,{\left( r^4 + R^4 \right) }^2\,\left( 5\,r^8 - R^4\,{r_0}^4 + 3\,r^4\,\left( R^4 - {r_0}^4 \right)  \right) 
     \Bigg) \,K_1(r) \cr
& + \Bigg( 3\,{k}^2\,r\,{\left( r^4 + R^4 \right) }^2\,
      \left( 5\,r^8 - R^4\,{r_0}^4 + 3\,r^4\,\left( R^4 - {r_0}^4 \right)  \right)  \cr
& \qquad{} - 
     32\,r^3\,\left( 3\,R^8\,{r_0}^4 + r^8\,\left( 6\,R^4 + 9\,{r_0}^4 \right)  + 
        r^4\,\left( -4\,R^8 + 2\,R^4\,{r_0}^4 \right)  \right)  \Bigg) \,L_1(r) \cr
& + 
  \left( r^4 - {r_0}^4 \right) \,\left( 9\,r^{12} + R^8\,{r_0}^4 + r^8\,\left( 34\,R^4 + 9\,{r_0}^4 \right)  + 
     5\,r^4\,\left( R^8 - 2\,R^4\,{r_0}^4 \right)  \right) \,A_1'(r) \cr
& - 
  \left( r^4 + R^4 \right) \,\left( 5\,r^4 - {r_0}^4 \right) \,
   \left( 5\,r^8 - R^4\,{r_0}^4 + 3\,r^4\,\left( R^4 - {r_0}^4 \right)  \right) \,K_1'(r) \cr
& - 
  3\,\left( r^4 + R^4 \right) \,\left( 5\,r^4 - {r_0}^4 \right) \,
   \left( 5\,r^8 - R^4\,{r_0}^4 + 3\,r^4\,\left( R^4 - {r_0}^4 \right)  \right) \,L_1'(r) \cr
& - 
  \left( r^4 + R^4 \right) \,\left( r^5 - r\,{r_0}^4 \right) \,
   \left( 5\,r^8 - R^4\,{r_0}^4 + 3\,r^4\,\left( R^4 - {r_0}^4 \right)  \right) \,K_1''(r) \cr
& - 
  3\,\left( r^4 + R^4 \right) \,\left( r^5 - r\,{r_0}^4 \right) \,
   \left( 5\,r^8 - R^4\,{r_0}^4 + 3\,r^4\,\left( R^4 - {r_0}^4 \right)  \right) \,L_1''(r) \,.
}

 \begin{figure}
  \centerline{\psfig{figure=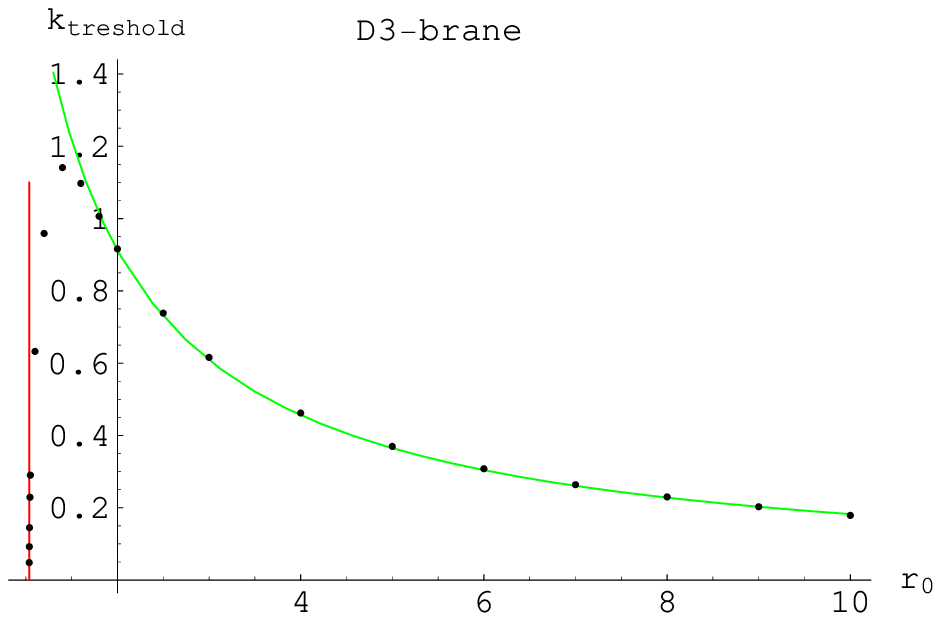,width=3in}}
  \centerline{\psfig{figure=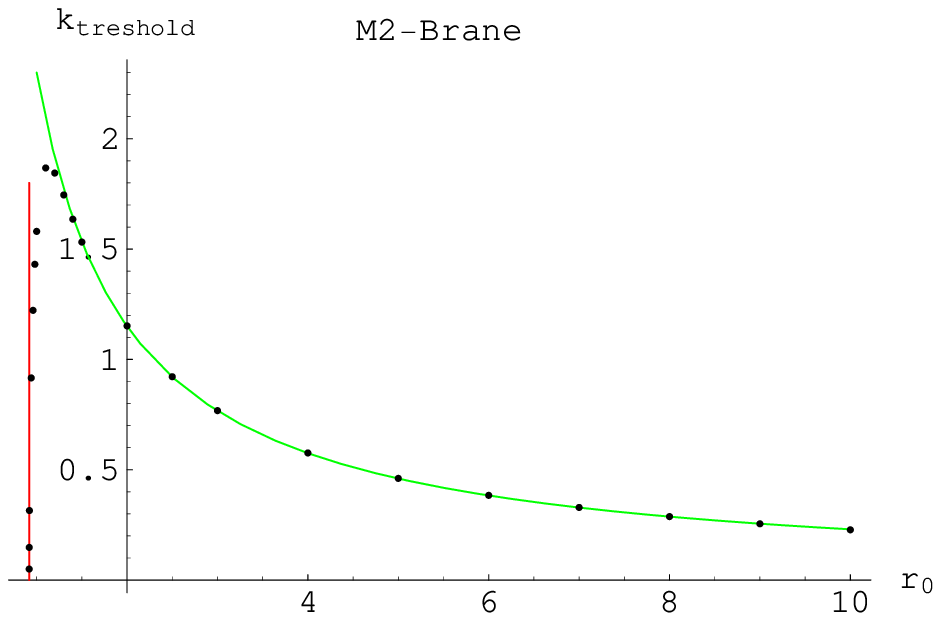,width=3in}}
  \centerline{\psfig{figure=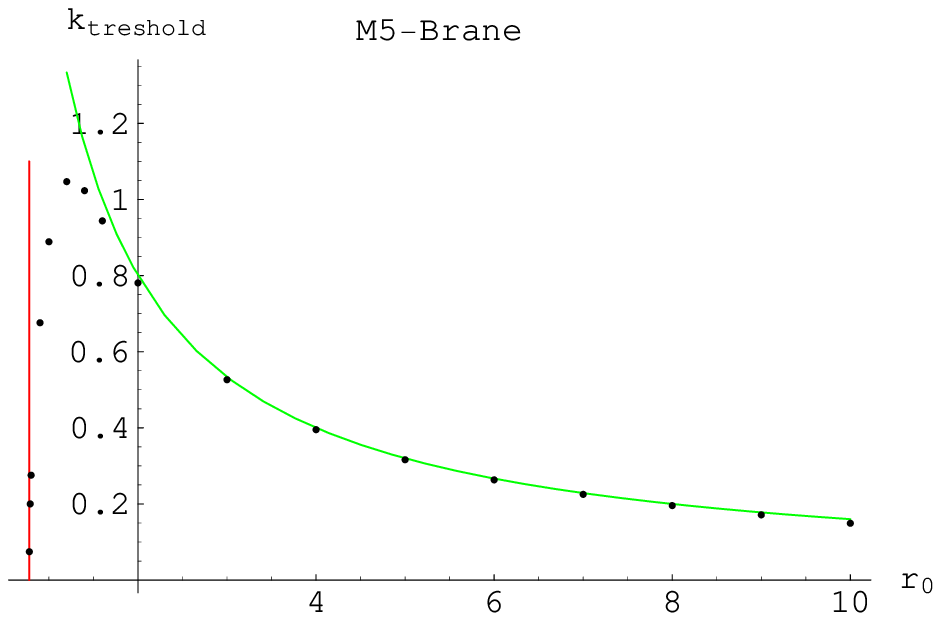,width=3in}}  
  \caption{Instabilities of black D3, M2, and M5-branes. 
     The dots represent threshold instabilities of the black branes 
     we found. The vertical lines indicate the thermodynamic 
     stability limits, which were found by \rzeroStr, with 
     $\tilde{R}=1$ for the D3-brane, $\tilde{R}=(1/3)^{1/6}$ for the
     M2-brane, and $\tilde{R}=(2/3)^{1/3}$ for the M5-brane.  
     The curves indicate 
     wavenumbers of threshold instabilities of {\it uncharged} black branes,
     with $r_0$ being the Schwarzchild radius.
     }\label{figD3M2M5}
 \end{figure}

Solving these linear ODE's numerically, one extracts using a shooting
algorithm the value of $k$ where a static perturbation exists, for any
specified non-extremal mass.  The results of this numerical
investigation are summarized in figure~\ref{figD3M2M5}. For large
non-extremality, the threshold wavenumbers fit nicely onto the curve
that gives those of an uncharged 3-brane. (This curve was obtained by
reading off the wavenumber of the threshold instability of an
uncharged 3-brane from the plots in \cite{glOne,glTwo}, and doing an
appropriate scaling to get the values for an arbitrary $r_0$.) As one
approaches the thermodynamic stability limit, the wavenumbers approach
to zero. In fact, by zooming into this region, we were able to show
that the wavenumbers go to zero approximately with a power-law
behavior, with the critical exponent being close to $1/2$: that is, $k
\sim \sqrt{r_0-r_0^*}$ where $r_0^*$ is the critical value where the
instability disappears.  Because $M$ is a non-singular function of
$r_0$, one could also write $k \sim \sqrt{M-M_*}$.
A naive way of understanding this is to think
of some effective theory describing the perturbations in terms of a
bosonic field with $m^2 < 0$.  Then the criterion for the
perturbations to be static is $\omega^2 = k^2 + m^2 = 0$, where $k$ is
the wave-number of a static perturbation.  The resulting equation, $k
= \sqrt{-m^2}$, predicts the observed scaling, provided $m^2 \sim
r_0^*-r_0$.  Such analytic behavior is a standard assumption: our
argument basically amounts to a naive application of Landau
theory.\footnote{It is actually a little too naive, because it
predicts an incorrect dispersion relation.  In \cite{glOne,glTwo}, it
was found that $\omega \to 0$ as $k \to 0$, whereas using $\omega^2 =
k^2 + m^2$ would suggest finite $\omega$ at $k=0$.  Possibly an
improved understanding could be based on hydrodynamic considerations.}

The numerical stability analyses for the M2-brane and M5-brane are
similar to the one for the D3-brane.  The results are summarized in
figure~\ref{figD3M2M5}. Once again, near the thermodynamic stability
limit, the wavenumbers go approximately as $\sqrt{r_0 - r_0^*}$.

\section{Conclusions}
\label{Conclude}

The stability of charged $p$-brane solutions was first attacked in
\cite{glTwo}.  Although the methods presented there are in principle
practicable for any solution, they require considerable computational
fortitude to apply.  We have argued that a simple notion of
universality classes is fairly broadly applicable to the question of
brane stability: one commonly finds a thermodynamic relation $S
\propto VT^\alpha$ for charged $p$-brane solutions in some limiting
regime of parameters, and when one does, the stability properties
depend only on $\alpha$---stability pertaining when $\alpha>0$.  Given
our treatment of the problem via a Kaluza-Klein reduction / truncation
to a 2+1-dimensional action involving scalars and gravity, the
dispersion relation of unstable modes would also appear to be the same
for any solution in the universality class labeled by a given
$\alpha$.

We have further shown, in section~\ref{Cross}, that when one
approaches a boundary of stability, the critical wavelength separating
stable from unstable modes diverges, and that it does so for the cases
studied with a critical exponent that could be guessed on the grounds
of a naive effective field theory argument.  It would be interesting
to see if other exponents arise from non-generic situations, for
instance a case where the temperature $T$ depends on mass as $T_{\rm
max}-T \sim |M-M_*|^\beta$ for some $\beta \ne 2$.  It is possible
that spinning brane solutions provide a venue for more intricate
thermodynamics \cite{cgOne,cgTwo}, but one would encounter there the
complication of chemical potentials for the various angular momenta.

For the $p$-branes of type~II string theory and M-theory, away from
extremality but without angular momenta, thermodynamic stability
pertains up to an upper mass limit given in equation~\Mratio.  For the
D3-brane, the M2-brane, and the M5-brane, our numerical analysis
supports the claim \cite{gmOne,gmTwo,Reall} that dynamical and
thermodynamic stability coincide.  Extending the analysis to other
$p$-branes should not be too difficult.  

In conclusion, it seems that the notion of universality classes that
we introduced in section~\ref{universality}, together with the
behavior at a boundary of stability explored in section~\ref{Cross},
represent a fairly comprehensive description of the qualitative
features of the Gregory-Laflamme instability in linearized
perturbation theory.  Of course, it would be desirable to go beyond
classical perturbation theory in understanding universality classes of
behaviors for non-uniform horizons.  Some of the methods of this paper
may prove useful in that broader context as well.

\section*{Acknowledgments}

We thank A.~Brandhuber, A.~Erickcek, Y.~Li, Y.-T.~Liu, and C.~Nunez
for useful discussions.  The work of S.S.G.\ is supported in part by
the Department of Energy under Grant No.\ DE-FG02-91ER40671.  The work
of A.O.\ is supported in part by the Department of Energy under Grant
No.\ DE-FG03-92ER40701.

\bibliography{univ}
\bibliographystyle{ssg}

\end{document}